\documentclass[12pt,reqno]{amsart}

\usepackage{bm}
\usepackage{amssymb}

\newtheorem{theorem}{Theorem}
\newtheorem{proposition}{Proposition}

\newtheorem{remark}{Remark}

\newtheorem{definition}{Definition}

\newcommand{\wconn}{\widetilde {\nabla}}
\newcommand{\conn}{\overline {\nabla}}

\newcommand{\wD}{\widetilde{D}}
\newcommand{\oD}{\overline{D}}

\newcommand{\R}{\mathbb{R}}
\newcommand{\s}{\mathbb{S}}

\newcommand{\T}{\mathcal{T}}

\newfont{\bb}{msbm10 at 12pt}

\def\pf{{\textit {Proof :} }}
\def\qed{\hfill{Q.E.D.}\smallskip}

\newcommand{\ls}{\setlength{\baselineskip}{12pt}
                 \setlength{\parskip}{3mm}}

\newcommand{\mysection}[1]{\section{#1}\setcounter{equation}{0}}

\newcommand{\bal}{\begin{align}}      \newcommand{\eal}{\end{align}}
\newcommand{\ba}{\begin{array}}      \newcommand{\ea}{\end{array}}
\newcommand{\bc}{\begin{center}}     \newcommand{\ec}{\end{center}}
\newcommand{\be}{\begin{enumerate}}  \newcommand{\ee}{\end{enumerate}}
\newcommand{\beq}{\begin{eqnarray}}  \newcommand{\eeq}{\end{eqnarray}}
\newcommand{\beQ}{\begin{eqnarray*}} \newcommand{\eeQ}{\end{eqnarray*}}
\newcommand{\bi}{\begin{itemize}}    \newcommand{\ei}{\end{itemize}}
\newcommand{\bt}{\begin{tabular}}    \newcommand{\et}{\end{tabular}}
\newcommand{\bdm}{\begin{displaymath}} \newcommand{\edm}{\end{displaymath}}

\begin{document}

\title{On A Quasi-local Mass}

\address{Institute of Mathematics,
Academy of Mathematics and Systems Science, Chinese Academy of
Sciences, Beijing 100190, China}

\author{Xiao Zhang}

\email{xzhang@amss.ac.cn}

\begin{abstract}
We modify previous quasi-local mass definition. The new definition
provides expressions of the quasi-local energy, the quasi-local
linear momentum and the quasi-local mass. And they are equal to the
ADM expressions at spatial infinity. Moreover, the new quasi-local
energy has the positivity property.
\end{abstract}



\maketitle

\mysection{Introduction} \ls

Recently, there are many attempts to define quasi-local mass
\cite{K, LY1, LY2, MST, WY1, ST, WY2, WY3, MTX} using certain idea
initiated by Brown and York \cite{BY1, BY2}. The definitions are
valid for spacelike 2-surfaces. The positivity property requires the
2-surfaces are spheres which are in certain initial data sets.

It is well-known that the Brown-York mass decreases for the round
sphere in time slices in the Schwarzschild spacetime. In order to
obtain a quasi-local mass of Brown-York's type with increasing
monotonicity property, we proposed a new definition by choosing
certain spinor norm as lapse function \cite{Z1,Z2}. But, as pointed
out in \cite{Z1}, our quasi-local mass are not equal to the ADM mass
at spatial infinity.

In this note, we modify our definition and provide expressions of
the quasi-local energy, the quasi-local linear momentum and the
quasi-local mass. In particular, we observe that the norm of
quasi-local linear momentum 3-vector is well-defined although this
3-vector can not be defined in a quasi-local sense. This is unlike
the ADM total linear momentum at spatial infinity, which both the
norm and the 3-vector are well-defined. We show also that these
quantities are equal to the corresponding ADM quantities at spatial
infinity. Moreover, the new modified quasi-local energy has the
positivity property. In particular, the vanishing of the quasi-local
mass implies the spacetime is flat along the 2-spheres, which is in
consequence of applying Brown and York's isometric embedding
approach. This later result does not seem to hold true in the other
recent definitions of quasi-local mass mentioned above.

The essential idea of our approach in this note as well as in
\cite{Z1, Z2} is to localize Witten's proof of the positive energy
theorem \cite{SY1, SY2, W, PT} to the (spacelike) bounded domain.
The similar approach was used by Dougan and Mason earlier to define
the quasi-local mass \cite{DM}. In our previous formulation, the
2-spheres are assumed to be apparent horizon only. In order that the
Dirac-Witten equation has a solution, we have to assume that the
mean curvature of spacelike hypersurfaces does not change sign. In
this note we include the case that at least one boundary 2-sphere
satisfies the strict inequality in apparent horizon conditions. This
will ensure the existence of the Dirac-Witten equation on any
spacelike hypersurfaces whose mean curvature may change sign, and
this existence theorem should be required in Dougan and Mason's
definition \cite{DM, S1, S2}.

Surprisingly, our quasi-local mass increases for the round sphere in
time slices in the Schwarzschild spacetime \cite{Z1} although we are
not able to prove the increasing monotonicity property in general
case. This should have deep physical reason which can be seen
naively as follows: In classical mechanics, the total mass is the
space integral of the local mass density. But this fact is no longer
true in general relativity as vacuum can have the total energy and
the total mass as well. However, Witten's energy formula \cite{W,
PT} indicates that the total energy and the total mass are the space
integrals of the ``generalized'' local mass density involving both
the energy-momentum tensors and spinors. From this point of view, we
think our approach is in a way toward the full understanding of
quasi-local quantities.

\mysection{Dirac-Witten equations} \ls

Let $(N, {\widetilde g})$ be a 4-dimensional spacetime which
satisfies the Einstein fields equations. Let $(M, g, p)$ be a smooth
initial data set. Fix a point $z \in M$ and an orthonormal basis $\{
e _{\alpha}\}$ of $T_z N$ with $e_0$ future-time-directed normal to
$M$ and $e_i$ tangent to $M$ ($1\leq i \leq 3$).

Denote by $\s $ the (local) spinor bundle of $N$. It exists globally
over $M$ and is called the hypersurface spinor bundle of $M$. Let
$\widetilde \nabla$ and $\conn$ be the Levi-Civita connections of
$\widetilde g$ and $g$ respectively,  the same symbols are used to
denote their lifts to the hypersurface spinor bundle. There exists a
Hermitian inner product $(\;,\;)$ on $\s$ along $M$ which is
compatible with the spin connection $\widetilde \nabla $. The
Clifford multiplication of any vector $\widetilde X$ of $N$ is
symmetric with respect to this inner product. However, this inner
product is not positive definite. As $e_0 ^2 =\mbox{id}$, $\s$ is
decomposed into the direct sums of two subbundles which are the
eigenbundles of $e_0$ along $M$. This indicates there exists a
positive definite Hermitian inner product defined by $\langle\;
,\;\rangle = (e _0 \cdot\; ,\;)$ on $\s$ along $M$.

The second fundamental form of the initial data set is defined as $p
_{ij}= \widetilde g (\widetilde \nabla _i e_0, e_j).$ Suppose $M$
has boundary $\Sigma $ which has finitely many connected components
$\Sigma ^1, \cdots, \Sigma ^l$, each of which is a topological
2-sphere and endowed with its induced Riemannian and spin
structures. Fix a point $z \in \Sigma$ and an orthonormal basis $\{
e _i\}$ of $T _z M$ with $e _r =e _1$ outward normal to $\Sigma $
and $e _a$ tangent to $\Sigma$ for $2 \leq a \leq 3$. Let $h _{ab} =
\langle \conn _a e _r, e _b \rangle$ be the second fundamental form
of $\Sigma $. Let $H = tr( h )$ be its mean curvature. $\Sigma $ is
a {\it future/past apparent horizon} if
 \begin{eqnarray}
H \mp tr (p | _{\Sigma }) \geq 0 \label{horizon}
 \end{eqnarray}
holds on $\Sigma $.

Denote by $\nabla $ the lift of the Levi-Civita connection of
$\Sigma $ to the spinor bundle $\s | _{\Sigma}$. The Dirac-Witten
operator along $M$, the Dirac operator of $M$ acting on $\s$ and the
Dirac operator of $\Sigma$ acting on $\s | _{\Sigma}$ are defined as
$ \widetilde D = e _i \cdot \widetilde \nabla _i$, $\oD = e _i \cdot
\conn _i$, $D=e _a \cdot \nabla _a$ respectively. Let $ P
_{\pm}=\frac{1}{2}(Id \pm e _0 \cdot e _r \cdot ) $ be the
projective operators on $\s | _{\Sigma}$. Then $\s | _{\Sigma} \cong
\s | _{\Sigma} ^+ \oplus \s | _{\Sigma} ^-$ is decomposed into the
direct sum of two equivalent complex 2-dimensional sub-bundles $\s |
_{\Sigma} ^\pm =\big\{\phi \in \s | _{\Sigma} : P_{\pm} \phi = \phi
\big\}.$

The following integral identity is given by the relation $
\widetilde \nabla _a =\nabla _a +\frac{1}{2}h _{ab} e _r \cdot e _b
\cdot -\frac{1}{2}p _{aj} e _0 \cdot e _j \cdot$ and the
Weitzenb{\"o}ck type formula for the Dirac-Witten operator
 \begin{eqnarray}
 \begin{aligned}
\int _M |\widetilde \nabla \phi |^2 &+ \langle \phi, \T \phi \rangle
- |\widetilde D \phi |^2
 = \int _{\Sigma} \langle \phi, (e _r\cdot D -\frac{H}{2})\phi
 \rangle\\
 &+\frac{1}{2}\int _{\Sigma }\langle \phi, tr (p | _{\Sigma }) e
_0 \cdot e _ r\cdot \phi \rangle -\frac{1}{2}\int _{\Sigma }\langle
\phi, p _{ar} e _0 \cdot e _a \cdot \phi\rangle \label{w2}
 \end{aligned}
 \end{eqnarray}
where $\T=\frac{1}{2}(T_{00}+T_{0i} e_0 \cdot e_i\cdot)$. Let $\phi
=\phi^+ +\phi ^-$. Since $e _0 \cdot e _r \cdot \phi^\pm =\pm
\phi^\pm$, we have $e _0 \cdot e _r \cdot (e _r\cdot D\phi^\pm )=\mp
e _r\cdot D\phi^\pm$ and $e _0 \cdot e _r \cdot (e _0 \cdot e _a
\cdot \phi^\pm) =\mp e _0 \cdot e _a \cdot \phi^\pm$, therefore the
right hand side of (\ref{w2}) is
 \begin{eqnarray*}
 \begin{aligned}
\mbox{RHS}=&\int _{\Sigma} \langle \phi^+, e _r\cdot D\phi^-
\rangle +\langle \phi^-, e _r\cdot D\phi^+ \rangle\\
&-\frac{1}{2}\int _{\Sigma }\big[H-tr (p | _{\Sigma })\big]
|\phi^+|^2 -\frac{1}{2}\int _{\Sigma }\big[H+tr (p |
_{\Sigma })\big] |\phi^-|^2 \\
&-\frac{1}{2}\int _{\Sigma }\langle \phi^+, p _{ar} e _0 \cdot e _a
\cdot \phi^-\rangle -\frac{1}{2}\int _{\Sigma }\langle \phi^-, p
_{ar} e _0 \cdot e _a \cdot \phi^+\rangle.
 \end{aligned}
 \end{eqnarray*}

Parts $(a)$ and $(b)$ of the following existences are proved in
\cite{Z1} which correspond to the case all $\Sigma_i$ are either
future or past apparent horizons. Part $(c)$ corresponds to the case
that at least one boundary 2-sphere satisfies the strict inequality
in apparent horizon conditions.
\begin{proposition}\label{prop1}
Let $(N, {\widetilde g})$ be a spacetime which satisfies the
dominant energy condition. Let $(M, g, p)$ be a smooth initial data
with the boundary $\Sigma$ which has finitely many multi-components
$\Sigma _i $, each of which is topological 2-sphere.

(a) If $tr _g (p) \geq 0$ and all $\Sigma_i$ are past apparent
horizons, then the following Dirac-Witten equation has a unique
smooth solution $\phi \in \Gamma (\s)$
 \begin{eqnarray}
 \left\{ \begin{array}{ccccc}
       \wD \phi &=&  0 & \mbox{in} & M\\
  P _{+} \phi &=& P _{+}  \phi _{0i} & \mbox{on} &\Sigma _{i} \\
    \end{array} \right . \label{existence1}
 \end{eqnarray}
for given $\phi _{0i} \in \Gamma (\s \big| _{\Sigma_i})$;

(b) If $tr _g (p) \leq 0$ and all $\Sigma_i$ are future apparent
horizons, then the following Dirac-Witten equation has a unique
smooth solution $\phi \in \Gamma (\s)$
 \begin{eqnarray}
 \left\{ \begin{array}{ccccc}
       \wD \phi &=&  0 & \mbox{in} & M\\
  P _{-} \phi &=& P _{-}  \phi _{0i} & \mbox{on} &\Sigma _{i} \\
    \end{array} \right . \label{existence2}
 \end{eqnarray}
for given $\phi _{0i} \in \Gamma (\s \big| _{\Sigma_i})$;

(c) If $\Sigma _i$, $1\leq i \leq k_0$, are past apparent horizons
and $\Sigma _j$, $k_0 +1 \leq j \leq l$, are future apparent
horizons, and there exists at least $l_0$, $1\leq l_0 \leq l$, such
that the strict inequality in (\ref{horizon}) holds for $l_0$, then
the following Dirac-Witten equation has a unique smooth solution
$\phi \in \Gamma (\s)$
 \begin{eqnarray}
 \left\{ \begin{array}{ccccc}
       \wD \phi &=&  0 & \mbox{in} & M\\
  P _{+} \phi &=& P _{+}  \phi _{0i} & \mbox{on} &\Sigma _{i} \\
  P _{-} \phi &=& P _{-}  \phi _{0j} & \mbox{on} &\Sigma _{j} \\
    \end{array} \right . \label{existence3}
 \end{eqnarray}
for given $\phi _{0i} \in \Gamma (\s \big| _{\Sigma_i})$, $\phi
_{0j} \in \Gamma (\s \big| _{\Sigma_j})$.
\end{proposition}
\pf To establish the existences of (\ref{existence1}),
(\ref{existence2}) and (\ref{existence3}), we only need to show that
any solution $\wD \phi =0$ is trivial when $\phi _{0i}=\phi _{0j}
=0$. We first study the case $(c)$, in this case (\ref{w2}) gives
 \begin{eqnarray*}
 \begin{aligned}
\int _M |\widetilde \nabla \phi |^2 + \langle \phi, \T \phi \rangle
 =&-\frac{1}{2}\sum _{1\leq i \leq k_0} \int _{\Sigma _i}\big[H+tr (p | _{\Sigma _i})\big]
|\phi^-|^2 \\
&-\frac{1}{2}\sum _{k_0+1\leq j \leq l}\int _{\Sigma _j }\big[H-tr
(p | _{\Sigma _j})\big] |\phi^+|^2.
 \end{aligned}
 \end{eqnarray*}
Since the strict inequality in (\ref{horizon}) holds for $l_0$, we
must have $\phi |_{\Sigma _{l_0}}=0$ and $\wconn _i \phi =0$ along
$M$. However, in cases $(a)$ and $(b)$, we can not conclude that
$\phi |_{\Sigma _{i}}=0$ on some $i$ directly from the apparent
horizon condition (\ref{horizon}). We need to use some other
argument. Since $\wD (e _0 \cdot \phi ) = -tr _g (p) \phi$, we have
 \begin{eqnarray*}
\sum _{1\leq i \leq l}\int _{\Sigma _i} \langle e _r \cdot \phi , e
_0 \cdot \phi \rangle = \int _M \langle \wD \phi, e _0 \cdot \phi
\rangle - \langle \phi, \wD (e _0 \cdot \phi)\rangle =\int _M tr _g
(p) | \phi | ^2.
 \end{eqnarray*}
If all $\Sigma _i$ are past apparent horizons and $tr _g (p) \geq
0$, or all $\Sigma _i$ are future apparent horizons and $tr _g (p)
\leq 0$, then the above equality gives that $\phi | _{\Sigma _i} =0$
for all $i$. We have also $\wconn _i \phi =0$ along $M$ in these
cases. Now $ |d | \phi | ^2| \leq 2 | \conn \phi | | \phi |\leq |p|
| \phi | ^2 \leq C_p | \phi | ^2, $ where $C_p =max |p|$. If there
exists a point $x_0 \in M$ such that $\phi (x _0) \neq 0$, then $$ |
\phi | (x) \geq | \phi | (x _0) e ^{-C _p \rho _{x _0} (x)}. $$ That
taking $x$ to the boundary $\Sigma _{l_0}$ will gives rise to a
contradiction. Therefore $\phi =0$ over $M$ and the existences of
(\ref{existence1}), (\ref{existence2}) and (\ref{existence3})
follow. \qed

\begin{proposition}\label{prop2}
Let $(N, {\widetilde g})$ be a spacetime which satisfies the
dominant energy condition. Let $(M, g, p)$ be a smooth initial data
with the boundary $\Sigma$ which has finitely many multi-components
$\Sigma _i $, each of which is topological 2-sphere. Suppose $\phi
$, $\psi$ are smooth spinors which satisfy $\wconn _i \phi =0$,
$\wconn _i \psi =0.$ Denote $\phi _0 ^\pm =P_\pm \phi |_{\Sigma
_i}$, $\psi _0 ^\pm =P_\pm \psi |_{\Sigma _i}$. If either $\phi _0
^+$, $\psi _0 ^+$ or $\phi _0 ^-$, $\psi _0 ^-$ are linearly
independent on $\Sigma _i$, then $\phi$, $\psi$ are linearly
independent in $M$.
\end{proposition}
\pf Let $C_1 \phi (x_0) +C_2 \psi (x_0)=0$ for complex constants
$C_1$, $C_2$ at some point $x_0 \in M$. We shall show that $C_1$,
$C_2$ must be zero. Denote $\Phi (x)=C_1 \phi (x) +C_2 \psi (x)$.
Let $\Omega$ be an open subset of $M$ where $\Phi$ is nonzero. Since
$\widetilde\nabla _i \Phi =0$, we know that $|\Phi(y)| \geq |\Phi
(x)| e ^{-C_p \rho (x,y)}$ for any $x, y \in \Omega$ where $C_p
=max|p|$. If $\Omega $ is nonempty, we can take $y$ to the boundary
of $\Omega$ or to the point $x_0$. This will give rise to a
contradiction. Therefore $\Phi (x) \equiv 0$. Hence $C_1 \phi _0
^\pm +C_2 \psi _0 ^\pm=0$ on $\Sigma _i$. Since $\phi _0 ^+$, $\psi
_0 ^+$ or $\phi _0 ^-$, $\psi _0 ^-$ is linearly independent on
$\Sigma _i$, we have $C_1=C_2=0$ and the proposition follows. \qed

\mysection{Quasi-local quantities} \ls

Let $(N, {\widetilde g})$ be a spacetime which satisfies the
dominant energy condition. Let $(M, g, p)$ be a smooth initial data
set with boundary $\Sigma $ which has finitely many connected
components $\Sigma _1, \cdots, \Sigma _l$, each of which is a
topological 2-sphere. Suppose that some $\Sigma _{i} $ can be
smoothly isometrically embedded into a smooth spacelike hypersurface
$\breve{M} ^3 $ in the Minkowski spacetime $\mathbb{R} ^{3,1}$ and
denote by $\aleph$ the isometric embedding. Let $\breve{\Sigma}
_{i}$ be the image of $\Sigma _{i}$ under the map $\aleph$. Let
$\breve{e}_r$ the unit vector outward normal to $\breve{\Sigma}
_{i}$ and $\breve{h}_{ij}$, $\breve{H}$ are the second fundamental
form, the mean curvature of $\breve{\Sigma} _{i}$ respectively.
Denote by $p_0 = \breve{p} \circ \aleph$, $H _0 = \breve{H} \circ
\aleph $ the pullbacks to $\Sigma _i$.

The isometric embedding $\aleph$ also induces an isometry between
the (intrinsic) spinor bundles of $\Sigma _{i}$ and $\breve{\Sigma}
_{i}$ together with their Dirac operators which are isomorphic to $e
_r \cdot D$ and $\breve{e} _r \cdot \breve{D}$ respectively. This
isometry can be extended to an isometry over the complex
2-dimensional sub-bundles of their hypersurface spinor bundles.
Denote by ${\breve{\s} ^{\breve{\Sigma} _{i}}}$ this sub-bundle of
$\breve{\s} | _{\breve{\Sigma} _{i}}$. We choose the same
orientations for $N$ and $\R^{3,1}$. Since $e_2 \cdot e_3 =\breve{e}
_2 \cdot \breve{e} _3$, $P_{\pm}$ is preserved by $\aleph $.
Therefore,
 \beq
\s | _{\Sigma _i} ^\pm \cong {\breve{\s}| ^\pm _{\breve{\Sigma}
_{i}}} \cong {\breve{\s} ^{\breve{\Sigma} _{i}}}. \label{s-pm}
 \eeq
Let $\breve{\phi}$ be a constant section of ${\breve{\s}
^{\breve{\Sigma} _{i}}}$ and denote $\phi _0 = \breve{\phi} \circ
\aleph $. Denote by $\breve{\Xi}$ the set of all these constant
spinors $\breve{\phi}$ with the unit norm. This set is isometric to
$S ^3$.

In this note, we introduce the following conditions on $M$:
 \bi
 \item[(i)] $tr _g(p) \geq 0$, $H | _{\Sigma _i} + tr (p | _{\Sigma
_i}) \geq 0$ for all $i$;\\
 \item[(ii)] $tr _g(p) \leq 0$, $H |
_{\Sigma _i} - tr (p | _{\Sigma _i}) \geq 0$ for all $i$;\\
 \item[(iii)] $H | _{\Sigma _i} + tr (p | _{\Sigma
_i}) \geq 0$ for $1\leq i \leq k_0$, $H | _{\Sigma _i} - tr (p |
_{\Sigma _i}) \geq 0$ for $k_0 +1 \leq j \leq l$, and there exists
at least $l_0$, $1\leq l_0 \leq l$, such that the strict inequality
holds for $l_0$.
 \ei
Now we provide the definition of quasi-local quantities for some
$\Sigma _{i}$ under these conditions. Recall \cite{Z1,Z2}

{\it Case 1.} If $\Sigma _{i}$ has positive Gauss curvature, then it
can be embedded smoothly isometrically into $\mathbb{R} ^3$ in the
Minkowski spacetime $\mathbb{R} ^{3,1}$. Denote by $\breve{\Sigma}
_{i}$ its image. In this case, $\breve{p}=0$.

Let $\phi $ be the unique solution of one of (\ref{existence1}),
(\ref{existence2}) and (\ref{existence3}) for some $\breve{\phi_i}
\in \breve{\Xi}$ and $\breve{\phi_j}=0$ for $j \neq i$. Denote
 \beQ
 \begin{aligned}
m(\Sigma _{i}, \breve{\phi})=&\frac{area(\Sigma _i)}{8\pi \int
_{\Sigma _i} |\phi|^2 } \Re \int _{\Sigma _{i}} \Big[(H _0 -H ) |
\phi | ^2 \\
&+ tr(p |_{\Sigma _{i}}) \langle \phi, e_0 \cdot e _r \cdot \phi
\rangle  - p _{ar} \langle \phi, e_0 \cdot e _a \cdot \phi \rangle
\Big].
 \end{aligned}
 \eeQ
Define
 \beQ
m^\flat _{i}=\min _{\breve{\Xi}}m(\Sigma _{i}, \breve{\phi}), \qquad
m^\natural _{i}=\max _{\breve{\Xi}}m(\Sigma _{i}, \breve{\phi}).
 \eeQ

{\it Case 2.} If $\Sigma _{i}$ has negative Gauss curvature $K
_{\Sigma _{i}} \geq -\kappa ^2$ ($\kappa > 0$) where $-\kappa ^2$ is
the minimum of the Gauss curvature. (Here we must choose the minimum
of the Gauss curvature instead of arbitrary lower bound, otherwise
the quasi-local mass defined in the following way might depend on
this arbitrary lower bound.) By \cite{P, CW}, $\Sigma _{i}$ can be
smoothly isometrically embedded into the hyperbolic space
$\mathbb{H} _{-\kappa ^2} ^3 $ with constant curvature $-\kappa ^2$
as a convex surface which bounds a convex domain in $\mathbb{H}
_{-\kappa ^2} ^3 $. Let $(t, x_1, x_2, x_3)$ be the spacetime
coordinates of $\mathbb{R} ^{3,1}$. Then $\mathbb{H} _{-\kappa ^2}
^3 $ is one-fold of the spacelike hypersurfaces $ \big\{ t^2 -x _1
^2 -x_2 ^2 -x _3 ^2 = \frac{1}{\kappa ^2} \big\}. $ The induced
metric of $\mathbb{H} _{-\kappa ^2} ^3 $ is
$
\breve{g} _{\mathbb{H} _{-\kappa ^2} ^3 } =\frac{dr^2}{1+ \kappa ^2
r ^2} +r ^2 (d \theta ^2 +\sin ^2 \theta d \psi ^2 )
$
It has the second fundamental form $\breve{p} ^{+} _{\mathbb{H}
_{-\kappa ^2} ^3 }= \kappa \breve{g} _{\mathbb{H} _{-\kappa ^2} ^3
}$ for the upper-fold $\{t
>0\}$ and $\breve{p} ^{-} _{\mathbb{H} _{-\kappa ^2} ^3 }= -\kappa \breve{g}
_{\mathbb{H} _{-\kappa ^2} ^3 }$ for the lower-fold $\{t <0\}$ with
respect to the future-time-directed normal.

If condition $(\mbox{i})$ holds, we embed $\Sigma _{i}$ into
upper-fold $\{t>0\}$. Since $\breve{\Sigma} _{i}$ is convex, we have
$\breve{H}+tr(\breve{p} | _{\breve{\Sigma}_{i}}) > 0.$ If condition
$(\mbox{ii})$ holds, we embed $\Sigma _{i}$ into lower-fold
$\{t<0\}$. We have $\breve{H}-tr(\breve{p} | _{\breve{\Sigma}_{i}})>
0$ in this case. If condition $(\mbox{iii})$ holds, we embed $\Sigma
_{i}$ into upper-fold if $H+tr(p | _{\Sigma_i})\geq 0$ and embed it into
lower-fold if $H-tr(p | _{\Sigma_i}) \geq 0$. Denote also by
$\breve{\Sigma} _{i}$ its image.

Let $\phi $ be the unique solution of one of (\ref{existence1}),
(\ref{existence2}) and (\ref{existence3}) for some $\breve{\phi_i}
\in \breve{\Xi}$ and $\breve{\phi_j}=0$ for $j \neq i$. Denote
 \beQ
 \begin{aligned}
m _{\pm}(\Sigma _{i}, \breve{\phi}) =& \frac{area(\Sigma _i)}{8\pi
\int _{\Sigma _i} |\phi|^2 } \Re \int _{\Sigma _{i}} \Big[(H _0 -H )
| \phi | ^2
 + tr(p |_{\Sigma _{i}})\langle \phi, e_0 \cdot e _r \cdot \phi
\rangle \\
&- p _{ar}\langle \phi, e_0 \cdot e _a \cdot \phi \rangle \Big] \mp
\frac{area(\Sigma _i) \kappa}{4\pi \int _{\Sigma _i} |\phi|^2}\int
_{\Sigma _{i}} \langle \phi, e_0 \cdot e _r \cdot \phi \rangle.
 \end{aligned}
 \eeQ
Define
 \beQ
 \begin{aligned}
m^\flat _{i}&=\left\{\begin{array}
                  {r@{:\;\;}l}
\min _{\breve{\Xi}}m _{+}(\Sigma _{i}, \breve{\phi}) & \mbox{if
$\Sigma _i$ is embedded into upper-fold}, \\
\min _{\breve{\Xi}}m _{-}(\Sigma _{i}, \breve{\phi}) & \mbox{if
$\Sigma _i$ is embedded into lower-fold}.
 \end{array} \right. \\
m^\natural _{i}&=\left\{\begin{array}
                  {r@{:\;\;}l}
\max _{\breve{\Xi}}m _{+}(\Sigma _{i}, \breve{\phi}) & \mbox{if
$\Sigma _i$ is embedded into upper-fold}, \\
\max _{\breve{\Xi}}m _{-}(\Sigma _{i}, \breve{\phi}) & \mbox{if
$\Sigma _i$ is embedded into lower-fold}.
 \end{array} \right.
 \end{aligned}
 \eeQ

{\it Case 3.} If $\Sigma _{i}$ has nonnegative Gauss curvature which
vanishes at some point, then it can be only $C^{1,1}$ embedded
isometrically into $\mathbb{R} ^3$ \cite{GL, HZ}. To avoid this
regularity problem, we first embed it smoothly isometrically into
hyperbolic space $\mathbb{H} _{-\kappa ^2} ^3 $ and then take the
limit $\kappa \rightarrow 0$. Define
 \beQ
 \begin{aligned}
m^\flat _{i}&=\left\{\begin{array}
                  {r@{:\;\;}l}
\lim _{\kappa \rightarrow 0}\min _{\breve{\Xi}}m _{+}(\Sigma _{i},
\breve{\phi}) & \mbox{if $\Sigma _i$ is embedded into upper-fold}, \\
\lim _{\kappa \rightarrow 0}\min _{\breve{\Xi}}m _{-}(\Sigma _{i},
\breve{\phi}) & \mbox{if $\Sigma _i$ is embedded into lower-fold}.
 \end{array} \right. \\
m^\natural _{i}&=\left\{\begin{array}
                  {r@{:\;\;}l}
\lim _{\kappa \rightarrow 0}\max _{\breve{\Xi}}m _{+}(\Sigma _{i},
\breve{\phi}) & \mbox{if $\Sigma _i$ is embedded into upper-fold}, \\
\lim _{\kappa \rightarrow 0}\max _{\breve{\Xi}}m _{-}(\Sigma _{i},
\breve{\phi}) & \mbox{if $\Sigma _i$ is embedded into lower-fold}.
 \end{array} \right.
 \end{aligned}
 \eeQ

\begin{remark}
The quasi-local quantities do not depend on the choice of spinors
after taking maximum and minimum on certain constant spinor space.
This procedure was not taken to rule out spinors in Dougan and
Mason's definition \cite{DM}.
\end{remark}

\begin{remark}
The solutions of (\ref{existence1}), (\ref{existence2}) and
(\ref{existence3}) for some $\breve{\phi_i} \in \breve{\Xi}$ and
$\breve{\phi_j}=0$ for $j \neq i$ are parameterized by the positive
or negative part of the constant Dirac spinors in
$\mathbb{R}^{3,1}$. So the set of these unit norm constant spinors
is $S^3$.
\end{remark}

\begin{remark} The normalized factor is omitted in the definitions of
$m(\Sigma _{i}, \breve{\phi})$ and $m _{\pm}(\Sigma _{i},
\breve{\phi})$ in \cite{Z1,Z2}.
\end{remark}

 \begin{definition}
Let $(N, {\widetilde g})$ be a spacetime which satisfies the
dominant energy condition. Let $(M, g, p)$ be a smooth initial data
set with the boundary $\Sigma$ which has finitely many
multi-components $\Sigma _i $, each of which is topological
2-sphere. If one of conditions $(\mbox{i})$, $(\mbox{ii})$,
$(\mbox{iii})$ holds, then the quasi-local energy $E^{loc} _{i}$,
the quasi-local linear momentum $|P^{loc}_{i}|$ and the quasi-local
mass $m ^{loc} _{i}$ of $\Sigma _{i}$ are defined respectively as
follows
 \beQ
E ^{loc} _{i} =\frac{m^\natural _{i}+ m^\flat _{i}}{2},\quad |P
^{loc} _{i}| =\frac{m^\natural _{i}- m^\flat _{i}}{2},\quad m ^{loc}
_{i} =\sqrt{m^\natural _{i} m^\flat_{i}}.
 \eeQ
 \end{definition}
From the following theorem, we know that $m^\natural _i\geq m^\flat
_i \geq 0$. Therefore these quasi-local quantities are well-defined.

It is straightforward that these quasi-local quantities all vanish
if $(N,{\widetilde g})$ is the Minkowski spacetime. On the other
hand, by the uniqueness, any solution of (\ref{existence1}) or
(\ref{existence2}) with boundary value in $\breve{\Xi}$ approaches a
constant spinor in asymptotically flat initial data sets. Therefore,
the limits at spatial infinity are
 \beQ
\lim _{r \rightarrow \infty} m^\natural _{i} =E +|P|,\quad \lim _{r
\rightarrow \infty} m^\flat _{i} =E -|P|
 \eeQ
where $E$, $P_k$ are the ADM total energy, the ADM total linear
momentum respectively and $|P|=\sqrt{P_1 ^2 +P_2 ^2 +P_3 ^2}$.
Denote by $m=\sqrt{E^2-P_1 ^2 -P_2 ^2 -P_3 ^2}$ the ADM mass. We
have
 \beQ
\lim _{r \rightarrow \infty} E^{loc} _{i} =E,\quad \lim _{r
\rightarrow \infty} |P ^{loc} _{i}| =|P|, \quad \lim _{r \rightarrow
\infty} m^{loc} _{i} =m.
 \eeQ

 \begin{theorem}\label{thm}
Let $(N, {\widetilde g})$ be a spacetime which satisfies the
dominant energy condition. Let $(M, g, p)$ be a smooth initial data
set with the boundary $\Sigma$ which has finitely many
multi-components $\Sigma _i $, each of which is topological
2-sphere. If one of conditions $(\mbox{i})$, $(\mbox{ii})$,
$(\mbox{iii})$ holds, then
 \begin{itemize}
 \item[(a)] $m^\flat _i
\geq 0 \Leftrightarrow E^{loc} _{i} \geq |P ^{loc} _{i}|$ for each
$i$;
 \item[(b)] If $E^{loc} _{i_0} =0$ for some $i_0$, then $N$ is
flat along $\Sigma _{i_0}$.
 \end{itemize}
 \end{theorem}
\pf Part (a) is a direct consequence of Proposition \ref{prop1} and
the following inequality proved in \cite{Z1,Z2}
 \beq
\int _{\Sigma _{i_0}} \langle \phi,e _r \cdot D \phi\rangle \leq
\frac{1}{2}\int _{\Sigma _{i_0}} \langle \phi, (H _0 -p _{0aa}e_0
\cdot e _r \cdot +p _{0ar} e_0 \cdot e _a \cdot )\phi \rangle.
\label{ineq}
 \eeq
For the completeness, we summarize the proof of (\ref{ineq}) here.
Since $\partial _i \breve{\phi}=0$, restricted on $\breve{\Sigma}
_{i_0}$, we have $\breve{\nabla} _a \breve{\phi}
+\frac{1}{2}\breve{h}
 _{ab} \breve{e} _r \cdot \breve{e} _b \cdot
\breve{\phi} -\frac{1}{2}\breve{p} _{aj} \breve{e} _0 \cdot
\breve{e} _j \cdot \breve{\phi} =0$. Pullback to $\Sigma _{i _0}$,
we obtain  $ e _r \cdot D \phi ^\pm _0  = \frac{H_0}{2} \phi ^\mp _0
\pm\frac{1}{2}p _{0aa} \phi ^\mp _0 +\frac{1}{2}p _{0ar} e _0\cdot e
_a \cdot \phi ^\pm _0. $ Using $ \int _{\Sigma _{i _0}} \langle \phi
^- _0, e _r \cdot D \phi ^+ _0 \rangle =\int _{\Sigma _{i _0}}
\langle e _r \cdot D\phi ^- _0, \phi ^+ _0\rangle $, we get
$
 \int _{\Sigma _{i _0}} (H _0 -p _{0aa} )|\phi _0 ^+ |
^2 =\int _{\Sigma _{i _0}}(H _0 +p _{0aa} ) |\phi _0 ^- | ^2.
$

The proof of (\ref{ineq}) as well as the theorem under condition
$(i)$ or condition $(iii)$ with $\Sigma _{i_0}$ the past apparent
horizon is the same as that under condition $(ii)$ or condition
$(iii)$ with $\Sigma _{i_0}$ the future apparent horizon. So we only
prove the first case. Let $\phi $ be the smooth solution of the
Dirac-Witten equation with $\phi ^+ =\phi _0 ^+$. We have
 \beQ
 \begin{aligned}
\int _{\Sigma _{i_0}} \langle \phi,e _r \cdot D \phi \rangle =& 2
\Re \int _{\Sigma _{i_0}} \langle \phi ^-,e _r \cdot D \phi _0
^+\rangle \\
=& \Re \int _{\Sigma _{i_0}} \langle \phi ^-, H _0 \phi _0 ^- + p
_{0aa} \phi _0 ^- +p _{0ar} e _0 \cdot e _r \cdot \phi _0 ^+ \rangle \\
\leq &\frac{1}{2}\int _{\Sigma _{i_0}} (H _0 +p _{0aa})(|\phi
^- | ^2+|\phi _0 ^- | ^2 )\\
 & + \Re \int _{\Sigma _{i_0}} \langle \phi ^-,p _{0ar} e _0\cdot e
_a \cdot \phi ^+ _0 \rangle \\
=& \frac{1}{2} \int _{\Sigma _{i_0}} (H _0 +p _{0aa} ) |\phi
^- | ^2 + (H _0 -p _{0aa} )|\phi _0 ^+ | ^2 \\
 &+\Re \int _{\Sigma _{i_0}} \langle \phi ^-,p _{0ar} e _0\cdot e
_a
\cdot \phi ^+ \rangle \\
=&\frac{1}{2} \int _{\Sigma _{i_0}} H _0 |\phi | ^2 +p _{0aa}
(|\phi ^-| ^2 -|\phi ^+ | ^2 ) \\
 &+\Re \int _{\Sigma _{i_0}} \langle \phi ^-,p _{0ar} e _0\cdot e
_a \cdot \phi ^+ \rangle \\
=&\frac{1}{2}\int _{\Sigma _{i_0}}
\langle \phi, (H _0 -p _{0aa}e_0 \cdot e _r \cdot +p _{0ar} e_0
\cdot e _a \cdot )\phi \rangle.
 \end{aligned}
 \eeQ

For part (b), if $E^{loc} _{i_0} =0$ for some $i_0$, then
$m^\natural _{i_0}=m^\flat _{i_0}=0$. Therefore, for any
$\breve{\phi} \in \breve{\Xi}$, $m(\Sigma _{i_0}, \breve{\phi})=0$.
This implies $\wconn _i \phi =0$. Let $\breve{\phi}_1,
\breve{\phi}_2 \in \breve{\Xi}$ be two linearly independent constant
spinors. Let $\phi _1 $, $\phi _2$ be the smooth solutions of one of
(\ref{existence1}), (\ref{existence2}) and (\ref{existence3})
corresponding to the boundary values $\breve{\phi}$, $\breve{\psi}$.
Then $\wconn _i \phi_1=\wconn _i \phi _2=0$. By Proposition
\ref{prop2}, $\phi _1$ and $\phi _2$ are linearly independent.
Choose the frame $e_\alpha$ such that $\wconn _\alpha e_\beta =0$.
Then
 \beq
-\frac{1}{4}\widetilde{R} _{ij \alpha \beta} e_\alpha e_\beta \phi
_a =[\wconn_i, \wconn_j ] \phi_a =0  \label{curvatures}
 \eeq
for $a =1,2$. Note that we are not able to obtain the vanishing of
$\widetilde{R} _{ij \alpha \beta}$ along $M$ as $\phi_a$ are complex
4-dimensional which might not belong to neither the positive
eigenbundle of $e_0$ nor the negative eigenbundle of $e_0$. However,
in the proof of (\ref{ineq}), the inequality $2\Re \langle \phi ^-,
\phi _0 ^- \rangle \leq |\phi ^- | ^2+|\phi _0 ^- | ^2$ is used on
$\Sigma _{i_0}$, and it is forced to be the equality when $E^{loc}
_{i_0} =0$. So $\phi ^- =\phi _0 ^-$ on $\Sigma _{i_0}$. Since the
boundary value condition gives $\phi ^+ =\phi _0 ^+$, we obtain
$\phi =\phi _0 $ on $\Sigma _{i_0}$. As $\phi_{0a}$ are pullbacks of
$\breve{\phi_{a}}$, $\phi _{0a}$ belong to the same eigenbundle of
$e_0$. This indicates the situation can be reduced to complex
2-dimensional spinor bundles on $\Sigma _{i_0}$ and
(\ref{curvatures}) holds true for two linearly independent constant
spinors $\phi _{0a}$ on $\Sigma _{i_0}$. Therefore $\widetilde{R}
_{ij \alpha \beta}=0$ on $\Sigma _{i_0}$. So $T_{00}=0$ on $\Sigma
_{i_0}$ and the dominant energy condition gives $T_{\alpha \beta}=0$
which implies $\widetilde{R} _{i0 \alpha \beta}=0$ on $\Sigma
_{i_0}$. Therefore the spacetime is flat along $\Sigma _{i_0}$. \qed

\mysection{Discussions}

In this note, we define the quasi-local quantities for a 2-sphere in
the spacelike hypersurface. For a spacelike 2-sphere $\Sigma$ in
spacetimes, we can take infimum of above defined quasi-local
quantities over the set of spacelike hypersurfaces enclosed by
$\Sigma$, if the set is nonempty. Because of Theorem \ref{thm}, the
infimum exists and is nonnegative. We can define this infimum as the
quasi-local quantities of $\Sigma$. If we think the choice of the
spacelike hypersurface for $\Sigma$ is gauge, we can call the {\it
infimum} gauge in this situation. Alternatively, if $\Sigma$ can
enclose a unique maximal hypersurface which satisfies $tr(p)=0$ in
spacetime (it is indeed the case in certain situation), then the
quasi-local quantities of $\Sigma$ can be defined in terms of the
geometry of this 2-sphere and the maximal hypersurface. And we can
call it the {\it maximal} gauge. The quasi-local quantities of
$\Sigma$ in these gauges can be understood not to depend on the
spacelike hypersurface.

\vspace{1cm}

{\footnotesize {\it Acknowledgements.} The author is indebted to N.
\'{O} Murchadha, Roh-Suan Tung and Naqing Xie for some valuable
conversations. This work is partially supported by NSF of China
(10421001, 10725105, 10731080), NKBRPC(2006CB805905) and the
Innovation Project of Chinese Academy of Sciences.}


\begin{thebibliography}{99}

\bibitem{BY1}
J.D. Brown and J.W. York, {\it Quasilocal energy in general
relativity}, Mathematical aspects of classical field theory
(Seattle, WA, 1991), 129-142, Contemp. Math., 132, Amer. Math. Soc.,
Providence, RI, 1992.

\bibitem{BY2}
J.D. Brown and J.W. York, {\it Quasilocal energy and conserved
charges derived from the gravitational action}, Phys. Rev. D(3) 47,
1407-1419 (1993).

\bibitem{CW}
M.P. do Carmo, F.W. Warner, {\it Rigidity and convexity of
hypersurfaces in spheres}, J. Diff. Geom. 4, 133-144 (1970).

\bibitem{DM}
A.J. Dougan and L.J. Mason, {\it Quasilocal mass constructions with
positive energy}, Phys. Rev. Lett. 67, 2119-2122 (1991).

\bibitem{GL}
F. Guan, Y.Y. Li, {\it The Weyl problem with nonnegative Gauss
curvature}, J. Diff. Geom., 39, 331-342 (1994).

\bibitem{HZ}
J.X. Hong, C. Zuily, {\it Isometric embedding of the 2-sphere with
nonnegative curvature in $R^3$}, Math. Z., 219, 323-334 (1995).

\bibitem{K}
J. Kijowski, {\it A simple derivation of canonical structure and
quasi-local Hamiltonians in general relativity}, Gen. Relat. Grav.
29, 307-343 (1997).

\bibitem{LY1}
C-C.M. Liu and S.T. Yau, {\it Positivity of quasilocal mass}, Phys.
Rev. Lett. 90, 231102 (2003).

\bibitem{LY2}
C-C.M. Liu and S.T. Yau, {\it Positivity of quasilocal mass II}, J.
Amer. Math. Soc. 19, 181 (2006).

\bibitem{MST}
N.\'{O} Murchadha, L.B. Szabados, K.P. Tod, {\it Comment on
``Positivity of quasilocal mass''}, Phys. Rev. Lett. 92, 259001
(2004).

\bibitem{MTX}
N.\'{O} Murchadha, R-S. Tung, N. Xie, {\it Quasi-local energy in
general relativity}, arXiv: 0905.0647 [gr-qc].

\bibitem{PT}
T. Parker and C.H. Taubes, {\it On Witten's proof of the positive
energy theorem}, Commun. Math. Phys. 84, 223-238 (1982).

\bibitem{P}
A.V. Pogorelov, {\it Some results on surface theory in the large},
Adv. Math. 1, 191-264 (1964).

\bibitem{SY1}
R. Schoen, S.T. Yau, {\it On the proof of the positive mass
conjecture in general relativity}, Commun. Math. Phys. 65, 45-76
(1979).

\bibitem{SY2}
R. Schoen, S.T. Yau, {\it Proof of the positive mass theorem. II},
Commun. Math. Phys. 79, 231-260 (1981).

\bibitem{ST}
Y. Shi, L-F. Tam, {\it Rigidity of compact manifolds and positivity
of quasi-local mass}, Class. Quantum Grav. 24 2357-2366 (2007).

\bibitem{S1}
L.B. Szabados, {\it On the positivity of the quasi-local mass},
Class. Quantum Grav. 10, 1899-1905 (1993).

\bibitem{S2}
L.B. Szabados, {\it Two-dimensional Sen connections and quasi-local
energy-momentum}, Class. Quantum Grav. 11, 1847-1866 (1994).


\bibitem{WY1}
M.T. Wang, S.T. Yau, {\it A generalization of Liu-Yau's quasi-local
mass}, Comm. Anal. Geom. 15, 249-282 (2007).

\bibitem{WY2}
M.T. Wang, S.T. Yau, {\it Quasi-local mass in general relativity},
Phys. Rev. Lett. 102, 021101 (2009).

\bibitem{WY3}
M.T. Wang, S.T. Yau, {\it Isometric embeddings into the Minkowski
space and new quasi-local mass}, Commun. Math. Phys. 288, 919¨C942
(2009).

\bibitem{W}
E. Witten, {\it A new proof of the positive energy
theorem}, Commun. Math. Phys. 80, 381-402 (1981).

\bibitem{Z1}
X. Zhang, {\it A new quasi-local mass and positivity}, Acta
Mathematica Sinica, English Series, 24(6), 881-890 (2008).

\bibitem{Z2}
X. Zhang, {\it A quasi-local mass for 2-spheres with negative Gauss
curvature}, Science in China Series A: Mathematics, 51, 1644-1650
(2008).
\end{thebibliography}
\end{document}